\documentstyle[12pt]{article}
\begin{document}
\title{Analysis of direct CP violation in $B^- \to
 D^0 D_s^-, D^0 D^-$ decays}
\author{A. K. Giri$^1$, R. Mohanta$^2$ and M. P. Khanna$^1$ \\
$^1$ {\it Physics Department, Panjab University, }\\
{\it Chandigarh - 160 014,
India}\\ $^2$ {\it School of Physics, University of Hyderabad,}\\
{\it Gachibowli, 
Hyderabad - 500 046, India }}
\maketitle
\begin{abstract}
We investigate the possibility of observing the direct CP
violation in the decay modes $B^- \to D^0 D_s^- $ and $D^0 D^-$
within the Standard Model. Including the contributions
arising from the tree, annihilation, QCD as well as
electroweak penguins with both time- and space-like components,
we find that the direct CP asymmetry in
 $B^- \to D^0 D_s^- $ is very small
$ \sim 0.2 $ \% but in $B^- \to D^0 D^- $
decay it can be as large as 4\%. Approximately $10^7 $
charged $B$ mesons are required to experimentally observe the
CP asymmetry parameter for the later case. Since this is easily accessible
with the currently running B factories,  the decay mode $B^- \to 
D^0 D^-$ may be pursued to look for CP violation.  \\
\\
PACS Numbers. : 11.30 Er, 13.25 Hw 
\end{abstract}
\section{Introduction}

CP violation is one of the least understood phenomena
in particle physics \cite{kog89,bigi00,silva99}, although it was
observed in $K^0 -\bar K^0 $ mixing system more than 35 years ago.
In the standard model (SM), CP violation arises from a complex phase 
in the Cabibbo-Kobayashi-Maskawa quark mixing matrix \cite{ref1}.
Outside the Kaon system, decays of $B$ mesons provide
rich ground for investigating $CP$ violation
\cite{buras98,quin98}. Within the SM, the $CP$ violation is
often characterized by the so-called unitarity triangle \cite{chau88}.
By measuring CP violating rate asymmetries in $B$ decays, one can extract
$\alpha$, $\beta$ and $\gamma$, the three interior angles of the unitarity
triangle. The sum of these three angles must be equal to
$180^\circ $ in the SM with three generations.
At present we are at the beginning
of the $B$-factory era in particle physics, which will provide us
valuable insights to understand the phenomena of CP violation.
One of the main
programmes of the presently running and the upcoming $B$ factories
is to measure the size of CP violation in as many $B$ decay modes
as possible so as to establish the pattern of CP violation in various $B$
decays. Among the most interesing $B$ decay channels, the ``gold plated"
mode $ B_d \to J/\psi K_s $, \cite{carter80} allows the
determination of the angle $\beta$ of the unitarity triangle of CKM
matrix. Recent measurement of CP asymmetry in the $ B^0 \to  J/\psi K^0 $
and other related processes e.g. $ \psi^\prime K^0,~ \eta_c K^0 $ etc by the
BELLE \cite{belle01} and BaBar \cite{babar01} detectors at the KEK and
SLAC $B$ factories together with the earlier measurement of CDF \cite{cdf00}
constitute the first significant signal of CP violation outside the
neutral Kaon system.

While the most promising proposal for
observing CP violation in the $B$-system involves the mixing between
neutral $B$ mesons \cite{kog89}, the decays  of charged $B$ mesons are also
of particular importance for establishing  the detailed nature
of CP violation. Since charged $B$ mesons cannot mix,
a measurement of the CP violating
observable in these decays would be a clear sign of ``direct
CP violation" which has been searched for in $K$-system for quite
long with indefinite success. Only recently, such kind of CP violating
effect has been observed in the $K$ system  by the
NA48 \cite{na48} and KTeV \cite{ktev00}
collaborations. For the bottom meson case usually the charmless rare
$B$ decay modes are preferred to study
the direct CP violation as these decay
modes proceed with more than one Feynman diagrams.
In this paper we would like to look for some additional decay
channels which could help us in establishing the presence of
CP violation as quickly as possible.  For this purpose we 
investigate the direct CP violating effecs
in the decays of charged $B$ mesons to two charmed mesons
i.e. $B^- \to D^0 D_s^- $ and $D^0 D^-$.
It is worth emphasizing that these decay modes are flavor self
tagging processes which should be favored for experimental reconstuctions.
The decay mode $B^- \to D^0 D_s^- $ has already been observed experimentally
with a branching ratio $(1.3 \pm 0.4) \%$ and the upper limit for
$B^- \to D^0 D^- $ channel is found to be $< 6.7 \times 10^{-3}$
\cite{pdg00}. These decay modes which are described by the
the quark level transitions as $ b \to c \bar c q $
($q=s/d$ for $D_s^-/D^-$ in
the final state) proceed through three distinct type of flavor
topologies. These are : the color allowed but
Cabibbo suppressed tree, annihilation and the QCD as well as electroweak
penguin diagrams. To get significant direct CP violation one would
require two interferring amplitudes of comparable strengths,
with different strong and weak phases.
The weak phases arise from the
superposition of various penguin contributions and the usual tree diagrams.
The strong phases are generated by the perturbative penguin loops
(hard final state interaction) \cite{soni79} or final state interactions
involving two different isospins. Since the decay modes we considered
here are single isospin channels i.e. the final states $D^0 D_s^-$ and
$D^0 D^-$ are with isospin $I=1/2$ and  1 respectively the second type
of FSI strong phase differences  are absent for these channels.
Therefore at the first sight it appears that direct CP
violating effects in these channels would be negligibly small as the
tree contribution dominates over the other diagrams and thus have been
overlooked in the literature. But detailed
calculation shows that it is indeed not so. In fact the CP violating
effects in $B^- \to D^0 D^- $ channel can be as large as few percent level
which can be experimentally accessible in the first round of $B$ factories.
The reason  for the existence of such a significant CP violating parameter
may be due to the fact that the tree diagram for $b \to c \bar c d $
transition is although colour allowed, it is doubly Cabibbo suppressed,
and hence its magnitude is not very much larger than the penguin
contributions. CP violating effects in the decays
of neutral $B$ meson into double charmed
mesons have been extensively studied in Refs. [16-19],
where it has been shown that, these channels can be used as
an alternative method to the  $J/\psi K_s$ mode
for the extraction of the angle $\beta$.

In our analysis,  we use the standard theoretical framework
to study the nonleptonic
$B^- \to D^0 D_s^-(D^-) $ decay modes,
which is based on the effective Hamiltonian
approach in conjuction with the factorization hypothesis.
The short distance QCD corrected Hamiltonian is 
calculated to next-to-leading order. The renormalization scheme and scale
problems with factorization approach for matrix elements can be circumvented
by employing the scale and scheme independent effective Wilson coefficients.
In the literature the contributions of
space-like penguins are neglected assuming form factor
suppression. But as pointed out in Ref. \cite{bcp6} the effect of space
like penguin amplitudes can be remarkably enhanced by the hadronic 
matrix elements involving $(V-A)(V+A)$ or $(S+P)(S-P) $ currents. 
Therefore we have included the 
space and time like contributions
of both QCD and EW penguins, the annihilation contribution
in addition to  the dominant  tree diagrams.
Assuming the factorization approximation, the matrix elements of
the tree and time-like penguin diagrams have been calculated in the BSW model
\cite{bcp7},
whereas for the evaluation of the matrix elements of the space and
annihilation diagrams we have employed the Lepage and Brodsky model
\cite{bcp8}.

The paper is organized as follows. In section II we briefly discuss
the effective Hamiltonian together with the quark level matrix elements
and the numerical value of the Wilson coefficients in the effective
Hamiltonian approach. Assuming the factorization
approximation, the matrix elements of tree and time-like
penguins are evaluated in the BSW model and
for the space-like and annihilation diagrams we use
LB (Lepage and Brodsky) model. Determination of the  CP violating asymmetry
is presented in section III
and section IV contains our conclusion. 

\section{Framework}

The effective Hamiltonian ${\cal H}_{eff} $ for the  decay modes
$ B^- \to D^0 D_s^- $ and $D^0 D^-$ which are described by the quark
level transitions $ b \to c \bar c q $ (where $q=s$ for the former and $d$
for the later) have three classes of flavour topologies : the dominant
tree, annihilation, and both QCD as well as electroweak penguins given by
\cite{buras98}
\begin{eqnarray}
{\cal H}_{eff}&= &\frac{G_F}{\sqrt 2} \biggr\{\
\lambda_u
[c_1(\mu) O_1^u(\mu)+c_2(\mu) O_2^u(\mu)]+
\lambda_c
[c_1(\mu) O_1^c(\mu)+c_2(\mu) O_2^c(\mu)] \nonumber\\
&+&(\lambda_u+\lambda_c)\sum_{i=3}^{10} c_i(\mu) O_i(\mu)\biggr\}
+{\rm h.c.}\;,
\end{eqnarray}
where $\lambda_u=V_{ub}V_{u q}^* $ and $\lambda_c=V_{cb}V_{c q}^* $
and $c_i(\mu ) $ are the Wilson coefficients evaluated at the
renormalization scale $\mu$. The four fermion operators $O_{1-10}
$ are given as

\begin{eqnarray}
&&O_1^u = (\bar u b)_{V-A}(\bar q u)_{V-A}\;,~~~~~~~~~~
O_2^u = (\bar u_\alpha b_\beta)_{V-A}(\bar q_\beta u_\alpha)_{V-A}\;,
\nonumber\\  
&&O_1^c = (\bar c b)_{V-A}(\bar q c)_{V-A}\;,~~~~~~~~~~
O_2^c = (\bar c_\alpha b_\beta)_{V-A}(\bar q_\beta c_\alpha)_{V-A}\;,
\nonumber\\
&&O_{3(5)} = (\bar q b)_{V-A}\sum_{q^\prime}(\bar q^\prime 
q^\prime)_{V-A(V+A)}\;,\nonumber\\
&&O_{4(6)} = (\bar q_\alpha b_\beta)_{V-A}\sum_{q^\prime}
(\bar q_\beta^\prime q_\alpha^\prime)_{V-A(V+A)}\;,
\nonumber\\
&&O_{7(9)} = \frac{3}{2}(\bar q b)_{V-A}\sum_{q^\prime}
e_{q^\prime}(\bar q^\prime 
q^\prime)_{V+A(V-A)}\;,\nonumber\\
&&O_{8(10)} = \frac{3}{2}(\bar q_\alpha b_\beta)_{V-A}\sum_{q^\prime}
e_{q^\prime}(\bar q_\beta^\prime q_\alpha^\prime)_{V+A(V-A)}\;,
\end{eqnarray}
where $O_{1,2}$ are the tree level current-current operators,
$O_{3-6}$  the QCD penguin operators and $O_{7-10} $  the EW penguin operators.
$(\bar q_1 q_2)_{(V\pm A)} $ denote the usual $(V \pm A) $ currents.
The sum over $q^\prime $  runs over the quark fields that are active
at the scale $\mu=O(m_b) $ i.e. $(q^\prime \in u,d,s,c,b)$. The
Wilson coefficients depend (in general) on the renormalization scheme
and the scale $\mu $ at which they are evaluated. In the next to leading 
order their values obtained in the naive dimensional regularization
(NDR) scheme at $\mu=m_b(m_b)$ as \cite{ref6}
$c_1=1.082, ~c_2=-0.185$,~$c_3=0.014, ~c_4=-0.035,~
c_5=0.009,~ c_6=-0.041$,~  $c_{7/\alpha}=-0.002,~ c_{8/\alpha}=0.054,~
c_{9/\alpha}=-1.292 $ and  $c_{10/\alpha} = 0.263 $.

However, the physical matrix elements $\langle P_1 P_2 |
{\cal H}_{eff} | B \rangle $ are obviously independent of
both the scheme and the scale. Hence the dependence on the
Wilson coefficients must be compensated by a comensurate
calculation of the hadronic matrix elements in a nonperturbative
framework such as lattice QCD. Presently, this is not a viable
strategy as the calculation of the matrix elements $\langle P_1 P_2 |
O_i | B \rangle $ is beyond the scope of the current lattice
technology. However, perturbation theory comes to (partial) rescue;
with the help of which one-loop matrix elements can be rewritten 
in terms of the operators and the effective Wilson coefficients
$c_i^{eff}$ which are scheme and scale independent :
\begin{equation}
\langle q  q^\prime \bar q^\prime |{\cal H}_{eff} |b \rangle=
\sum_{i,j}c_i^{eff}(\mu)  
\langle q  q^\prime \bar q^\prime |O_j |b \rangle^{tree}\;.
\end{equation}
The effective Wilson coefficients $c_i^{eff}(\mu)$ may be expressed as
\cite{ref7}

\begin{eqnarray}
&&c_1^{eff}|_{\mu=m_b}=c_1(\mu)+\frac{\alpha_s}{4\pi} \left ( \gamma^{(0)T} 
\ln \frac{m_b}{\mu}+\hat{r}^T \right )_{1i}c_i(\mu)\;,\nonumber \\
&&c_2^{eff}|_{\mu=m_b}=c_2(\mu)+\frac{\alpha_s}{4\pi} \left ( \gamma^{(0)T} 
\ln \frac{m_b}{\mu}+\hat r^T \right )_{2i}c_i(\mu)\;,\nonumber \\
&&c_3^{eff}|_{\mu=m_b}=c_3(\mu)+\frac{\alpha_s}{4\pi} \left ( \gamma^{(0)T} 
\ln \frac{m_b}{\mu}+\hat r^T \right )_{3i}c_i(\mu)-\frac{\alpha_s}
{24 \pi}(C_t+C_p+C_g)\;,\nonumber \\
&&c_4^{eff}|_{\mu=m_b}=c_4(\mu)+\frac{\alpha_s}{4\pi} \left ( \gamma^{(0)T} 
\ln \frac{m_b}{\mu}+\hat r^T \right )_{4i}c_i(\mu)
+\frac{\alpha_s}{8 \pi}(C_t+C_p+C_g)\;,\nonumber \\
&&c_5^{eff}|_{\mu=m_b}=c_5(\mu)+\frac{\alpha_s}{4\pi} \left ( \gamma^{(0)T} 
\ln \frac{m_b}{\mu}+\hat r^T \right )_{5i}c_i(\mu)
-\frac{\alpha_s}{24 \pi}(C_t+C_p+C_g)\;,\nonumber \\
&&c_6^{eff}|_{\mu=m_b}=c_6(\mu)+\frac{\alpha_s}{4\pi} \left ( \gamma^{(0)T} 
\ln \frac{m_b}{\mu}+\hat r^T \right )_{6i}c_i(\mu)
+\frac{\alpha_s}{8 \pi}(C_t+C_p+C_g)\;,\nonumber \\
&&c_7^{eff}|_{\mu=m_b}=c_7(\mu)+\frac{\alpha_s}{4\pi} \left ( \gamma^{(0)T} 
\ln \frac{m_b}{\mu}+\hat r^T \right )_{7i}c_i(\mu)
+\frac{\alpha}{8 \pi}C_e\;,\nonumber \\
&&c_8^{eff}|_{\mu=m_b}=c_8(\mu)+\frac{\alpha_s}{4\pi} \left ( \gamma^{(0)T} 
\ln \frac{m_b}{\mu}+\hat r^T \right )_{8i}c_i(\mu)\;,\nonumber \\
&&c_9^{eff}|_{\mu=m_b}=c_9(\mu)+\frac{\alpha_s}{4\pi} \left ( \gamma^{(0)T} 
\ln \frac{m_b}{\mu}+\hat r^T \right )_{9i}c_i(\mu)
+\frac{\alpha}{8 \pi} C_e\;,\nonumber \\
&&c_{10}^{eff}|_{\mu=m_b}=c_{10}(\mu)+\frac{\alpha_s}{4\pi}
\left ( \gamma^{(0)T} 
\ln \frac{m_b}{\mu}+\hat r^T \right )_{10i}c_i(\mu)\;.
\end{eqnarray}
where $\hat r^T $ and $\gamma^{(0) T} $, the transpose of the matrices 
$\hat r $ and $\gamma^{(0)} $, arise from the vertex corrections to the
operators $O_1-O_{10} $ derived in \cite{ref8}, which are
explicitly given in Ref. \cite{ref12}

The quantities $C_t $, $C_p$, $C_e$ and $C_g$ are arising
from the peguin type
diagrams of the operators $O_{1,2}$, the QCD penguin type
diagrams of the operators $O_{3}-O_6$, the electroweak penguin
type diagrams of $O_{1,2}$ and the tree level diagrams of
the dipole operator $O_g$ respectively, which
are given in the NDR scheme
(after ${\overline {\rm MS}} $ renormalization) by
 
\begin{eqnarray}
&&C_t=-\left ( \frac{\lambda_u}{\lambda_t} \tilde{G}(m_u)
+\frac{\lambda_c}{\lambda_t} \tilde G(m_c) \right ) c_1\nonumber\\
&&C_p=[\tilde G(m_s)+\tilde G(m_b)]c_3+\sum_{i=u,d,s,c,b}\tilde G(m_i)
(c_4+c_6)\nonumber\\
&&C_g=-\frac{2m_b}{\sqrt {\langle k^2 \rangle} } c_g^{eff}\;,
~~~~~~~~~c_g^{eff}=-1.043 \nonumber\\
&&C_e=-\frac{8}{9}\left ( \frac{\lambda_u}{\lambda_t} \tilde{G}(m_u)
+\frac{\lambda_c}{\lambda_t} \tilde G(m_c) \right )( c_1+3c_2)\nonumber\\
&&\tilde G(m_q)=\frac{2}{3}-G(m_q,k,\mu)
\end{eqnarray}

\begin{equation}
G(m,k,\mu)=-4 \int_0^1 dx ~x(1-x)\ln \left ( \frac{m^2-k^2x(1-x)}
{\mu^2} \right )\;,
\end{equation}

It should be noted that the quantities $C_t$, $C_p $ $C_e$ and $C_g$
depend on the CKM matrix elements, the quark masses, the scale $\mu $ and
$k^2 $, the momentum transferred by the virtual particles apearing in the
penguin diagrams. In the factorization approximation there is no
model independent way to keep track of the $k^2 $ dependence; the actual
value of $k^2$ is model dependent. From simple kinematics \cite{ref9}
one expects $k^2 $ to be typically in the range
\begin{equation}
\frac{m_b^2}{4} \leq k^2 \leq \frac{m_b^2}{2}\;.
\end{equation} 
Since the branching ratio and the CP asymmetry depend crucially
on the parameter $k^2$, here we would like to take a specific value
for it based on valence quark approximation instead 
of the conventionally used value $k^2=m_b^2/2 $. As discussed in
Ref. \cite{bcp6} the averaged value of the squared momentum transfer
for $B^-(b \bar u ) \to D^0(c \bar u) D_q^-(q \bar c) $ is given as

\begin{equation}
\langle k^2 \rangle = m_b^2+m_q^2-2m_b E_q \label{eq:bc1}
\end{equation}
where the energy of the quark $q$ in the final $D_q^- $ particle is
determinable from
\begin{equation}
E_q+\sqrt{E_q^2-m_q^2+m_c^2}+\sqrt{4E_q^2-4m_q^2+m_c^2}=m_b\label{eq:eqn1}
\end{equation}
for time-like penguin channels; or from
  
\begin{equation}
E_q+\sqrt{E_q^2-m_q^2+m_u^2}=m_b+m_u\label{eq:bc2}
\end{equation}
for space-like penguin diagrams. 
$m_b$, $m_q$ and $m_c$ denote the masses of the decaying
$b$-quark, daughter $q$-quark and the $c$-quark (created as $c \bar c$ pair
from the virtual gluon, photon or Z-particle in the penguin loop). 
For numerical calculations, we have taken the CKM matrix elements
expressed in terms of the Wolfenstein parameters with values
$A=0.815 $, $\lambda= \sin \theta_c $=0.2205, $\rho=0.175 $
and $\eta=0.37 $ \cite{ref12}. The choice of $\rho $ and 
$\eta $ correspond to the 
CKM triangle : $\alpha = 91^{\circ} $, $\beta = 24^{\circ}$ and 
$\gamma = 65^{\circ}$.
At scale $\mu \sim m_b$, we use the current quark masses as \cite{ref12}
$m_u(m_b)$ = 3.2 MeV, $ m_d (m_b)$ = 6.4 MeV, $m_s(m_b)$ = 90 MeV,
$m_c (m_b)$= 0.95 GeV and $m_b (m_b)$= 4.34 GeV. With
the specific value of $k^2$ obtained from Eqns. (8-10), we 
obtain the values of the effective
renormalization scheme and scale independent Wilson coefficients
for $b \to s $  and $b \to d$ transitions as given in Table-1. 

Now we want to calculate the matrix element $ \langle D_q^- D^0|
O_i | B^- \rangle $ using the factorization approximation, 
where $O_i$ are the four quark current operators listed above.
In this approximation, the hadronic matrix elements of the 
four quark operators $(\bar c b)_{(V-A)}(\bar q c)_{(V-A)}$ 
split into the product of two matrix elements, 
$\langle D^0 |(\bar c  b)_{(V-A)} | B^- \rangle $
and $\langle D_q^- | (\bar q c)_{(V-A)} | 0 \rangle $ where 
Fierz transformation has been used so that flavor 
quantum numbers of the currents match
with those of the hadrons. Since Fierz rearranging yields operators which are
in the color singlet-singlet and octet-octet forms, this procedure results, 
in general, in matrix elements which have the right flavor quantum
numbers but involve both singlet-singlet and octet-octet 
current operators. However, there is no experimental information available
for the octet-octet part. So in the factorization approximation,
one discards the color octet-octet piece and compensates this by
treating $N_c$, the numbers of colors as a free parameter, and
its value is extracted from the data of two body nonleptonic decays.

The matrix elements of the $(V-A)(V+A)$ operators i.e. ($O_6~\&~
O_8 $) can be transformed into $(V-A)(V-A)$ form by using Fierz ordering and 
the Dirac equation, which are given as
\begin{equation}
\langle D_q^- D^0 |O_6 |B^- \rangle=R_q
\langle D_q^- D^0 |O_4 |B^- \rangle
\end{equation}
with
\begin{equation}
R_q=\frac{2 m_{D_q^-}}{(m_b-m_c)(m_q+m_c)}\;,
\end{equation}
where the quark masses are the current quark masses. The same
relation works for $O_8$.

Hence, one obtains the transition amplitude for $B^- \to D_s^- D^0 $
and $D^-D^0 $ as
(where the factor $G_F/\sqrt 2 $ is suppressed)

\begin{eqnarray}
A(B^- & \to & D_s^-  D^0)=\lambda_u \biggr \{
\biggr(a_4+a_{10}+(a_6+a_8)R_s \biggr)X^{(BD^0, D_s^-)}
\nonumber\\
&+&\biggr(a_1+a_4+a_{10}+(a_6+a_8)R_s^\prime \biggr)X^{(B,D^0 D_s^-)}
\biggr\}\nonumber\\
&+&\lambda_c \biggr\{ \biggr(a_1+a_4+a_{10}+(a_6+
a_8) R_s\biggr)X^{(BD^0, D_s^-)}\nonumber\\
&+&\biggr(a_4+a_{10}+(a_6+a_8)R_s^\prime \biggr)X^{(B,D^0 D_s^-)}
\biggr\}\label{eq:bqn3}
\end{eqnarray}

\begin{eqnarray}
A(B^- & \to & D^- \bar D^0)=\lambda_u \biggr\{ \biggr(a_4+
a_{10}+(a_6+a_8)R_d \biggr)X^{(BD^0, D^-)}\nonumber\\
&+& \biggr(a_1+a_4+a_{10}+(a_6+a_8)R_d^\prime \biggr)X^{(B,D^0 D^-)}
\biggr\}\nonumber\\
&+&\lambda_c \biggr\{ \biggr(a_1+a_4+a_{10}+(a_6+
a_8) R_d \biggr)X^{(BD^0, D^-)}\nonumber\\
&+& \biggr(a_4+a_{10}+(a_6+a_8)R_d^\prime \biggr)X^{(B,D^0 D^-)}
\biggr\}
\end{eqnarray}
where
\begin{eqnarray}
X^{(BD^0, D_q^-)}&=&\langle D_s^- | (\bar q c)|0\rangle \langle D^0
|(\bar c b)|
B \rangle\;\nonumber\\
X^{(B, D_q^- D^0)}&=&
\langle D^0 D_q^- |(\bar q c)|
0 \rangle\ \langle 0 | (\bar u b)|B\rangle\;.
\end{eqnarray}
$X^{(BD^0, D_q^-)}$ denotes matrix elements of the tree
and time like penguins where as
$X^{(B,D^0 D_q^-)}$ stand for the annihilation and space-like
amplitudes.
\begin{equation}
R_{q^\prime}=\frac{2 m_B^2}{(m_q-m_u)(m_b+m_u)}\;,
\end{equation}
arises from the transformation of $(V-A)(V+A)$ operators into $(V-A)(V-A)$
form for space-like penguins. It should be noted that
$\lambda_u =V_{ub}V_{us}^* $ for $B^- \to D^0 D_s^-$
whereas $\lambda_u =V_{ub}V_{ud}^* $ for $B^- \to D^0 D^-$ and similar
expressions for $\lambda_c$.

The coefficients $a_1,~a_2 \cdots~a_{10} $ are combinations of the
effective Wilson coefficients given as
\begin{equation}
a_{2i -1}=c_{2i-1}^{eff}+\frac{1}{N_c^{eff}}c_{2i}^{eff}~~~
a_{2i}=c_{2i}^{eff}+\frac{1}{N_c^{eff}}c_{2i-1}^{eff}~~~~
i=1,2 \cdots 5\;,
\end{equation}
where $N_c^{eff}$ is the effective number of colors treated as free
parameter in order to  model the nonfactorizable contributions
to the matrix elements and its value can be extracted from the
two body nonleptonic $B$ decays. A recent analysis of $B \to D \pi $ data
gives $N_c^{eff} \sim 2 $ \cite{ref11}.
Therefore, in our analysis, we take two
sets of values for $N_c^{eff} $ i.e.,
$N_c^{eff}=2 $ and $N_c^{eff}=3 $, which characterizes naive factorization.

The factorized hadronic matrix elements are evaluated using the 
BSW model \cite{bcp7}, which are given as
\begin{equation}
X^{(B D^0,D_q^- )}= i f_{D_q} F_0^{BD}(m_{D_q}^2)
(m_B^2 -m_{D^0}^2)
\end{equation}

The matrix element of the annihilation and space-like penguins
are given as \cite{bcp6}
\begin{equation}
\langle D^0 D_q^- |(\bar q u)(\bar u b) |B^- \rangle
=i f_B f_+^a(m_B^2) \left [m_{D_q}^2-m_{D^0}^2 -
\frac{m_{D_q} -m_{D^0}}{m_{D_q}+m_{D^0}} m_B^2 \right ]\;,\label{eq:eqn5}
\end{equation}
where the value of the annihilation form factor is given as
$f_+^a(m_B^2)=i 16 \pi \alpha_s f_B^2/m_B^2 $ \cite{bcp8}.

After obtaining the transition amplitude, the branching ratio is
given as
\begin{equation}
BR=\frac{|{\bf p}|}{8 \pi m_B^2} \frac{|A(B^- \to D^0 D_q^-)|^2}
{\Gamma}\;,\label{eq:eqn4}
\end{equation}
where $|{\bf p}| $ is the momentum of the emitted particles and $\Gamma $
is the total decay width.

Using eqns (\ref{eq:bqn3})-(\ref{eq:eqn5}) we obtain the transition
amplitude (in the unit of $G_F/{\sqrt 2}  $) as
\begin{eqnarray}
A(B^- &\to & D^0 D_s^-)=  \lambda_u(0.1898-i0.6483 )
+\lambda_c(0.1889+i4.418)
\nonumber\\
&& [\lambda_u(0.2019-i0.6817 )
+\lambda_c(0.201+i 4.698)]\;,
\end{eqnarray}

\begin{eqnarray}
A(B^- &\to & D^0 D^-)=  \lambda_u(0.2259-i0.5616 )
+\lambda_c(0.2259+i4.8185)
\nonumber\\
&& [\lambda_u(0.2393-i0.589 )
+\lambda_c(0.2393+i 5.124)]\;,
\end{eqnarray}
where we have used the decay constants (in MeV) as $f_{D_s}=280 $,
$f_{D}=300 $ \cite{pdg00} and $f_B$=180
\cite{cs98}. In the above equations,
the upper values correspond to $N_c^{eff}=2 $
and the lower bracketed values to $N_c^{eff}=3 $.

\section{CP Violating Asymmetry}

For charge $B^{\mp}$ decays, the CP violating
rate asymmetries in partial decay rates are defined as 

\begin{equation}
a_{cp} =\frac{\Gamma (B^- \to f^-) - \Gamma (B^+ \to f^+)}
{\Gamma (B^- \to f^-) + \Gamma ( B^+ \to f^+)}\;.
\end{equation}
As these decays are all self tagging the measurement of these CP
violating asymmetry is essentially a counting experiment in well
defined final states. Their rate asymmetries require both weak and
strong phase differences in interfereing amplitudes. The weak phase
difference arises from the superposition of amplitudes from various
tree (current-current) and penguin diagrams. The strong phase which
are needed to obtain nonzero values for $a_{cp}$ are generated by
absorptive parts in penguin diagrams (hard final state interactions).

For the $B$ meson decaying  to a final state $f$ and the charge conjugated
$ B^-\to f$ we may, without any loss of generality, write the transition
amplitude as
\begin{equation}
A(f) =\lambda_u A_u e^{i\delta_u} + \lambda_c A_c e^{i\delta_c}
\end{equation}
\begin{equation}
\bar A(f) =\lambda_u^* A_u e^{i\delta_u} + \lambda_c^* A_c e^{i\delta_c}
\end{equation}
where $\lambda_i = V_{ib} V_{iq}^*$, $A_u$ and $A_c$ denote the
contribution from penguin operators proportional to the product of
CKM matrix elements $\lambda_u$ and $\lambda_c$ respectively. The
corresponding strong phases are denoted by $\delta_u$ and $\delta_c$
respectively.

Thus the direct CP violating asymmetry is given as,
\begin{eqnarray}
a_{cp} &=&\frac{-2~ {\rm Im}(\lambda_u \lambda_c^*)~ {\rm Im}
(A_u A_c^*)}
{|\lambda_u A_u|^2 +|\lambda_c A_c|^2 + 2~{\rm Re}(\lambda_u \lambda_c^*)
~{\rm Re}(A_u A_c^*)}\nonumber\\
&&\nonumber\\
&=& \frac{2 \sin \gamma \sin (\delta_u -\delta_c)}
{|\frac{\lambda_u A_u}{\lambda_c A_c}| + |\frac{\lambda_c A_c}
{\lambda_u A_u}| + 2\cos \gamma \cos (\delta_u -\delta_c)}\;,
\label{eq:eqn7}
\end{eqnarray}
where the weak phases entering in the $b \to s/d $ transition is equal to
$(-\gamma )$, as we are using Wolfenstein approximation in which
$\lambda_c$ has no weak phase and the phase of $\lambda_u $
is $-\gamma $.
The strong phase $(\delta_u -\delta_c)$ is caused by the final state
interactions.

The strong phases are given by,

\begin{equation}
\sin (\delta_u - \delta_c) =\frac{1}{|A_u A_c|} ({\rm Im} A_u~ {\rm Re}
A_c
-{\rm Im} A_c~ {\rm Re} A_u)
\end{equation}
\begin{equation}
\cos (\delta_u - \delta_c) =\frac{1}{|A_c A_u|} ( {\rm Re} A_u~
{\rm Re} A_c
+ {\rm Im} A_u ~{\rm Im} A_c)
\end{equation}

\section{Conclusion}

Using the next-to-leading order QCD corrected effective Hamiltonian,
the scale and scheme independent Wilson coefficients, we have systematically
studied the two charm hadronic decay modes $B^- \to D^0 D_s^- $ and $D^0
D^-$
within the framework of generalized factorization. The nonfactorizable
contributions are parametrized in terms of $N_c^{eff} $,
the effective number of colors.
For numerical calculations, we have used two different sets of values for
this parameters: (i) $N_c^{eff}=2$, (ii)
$N_c^{eff}=3$ which holds for naive factorization. The existence of a
direct CP
violating rate asymmetry  requires two interferring amplitudes
having different
CP nonconserving weak phases and CP conserving strong phases.
The former may arise either from the Standard Model CKM matrix
or from new physics while the latter may arise from the
absorptive part of a penguin diagram or from final state interaction
effects of two different isospins. Since the channels we considered here
are single isospin channels, the second class of strong phase differences do
not arise for these channels. In our analysis, the weak phases are due to the CKM matrix
and the strong phase differences  arise due to absorptive part of penguin
diagrams. The branching ratio and the CP violating
asymmetry parameter are estimated 
using Eqs.(\ref{eq:eqn4}) and (\ref{eq:eqn7}) 
and are presented in Table-2

From the results we have observed the following:

1. The predicted branching ratio for the decay mode $B^- \to D^0 D_s^-$
agrees very well with the experimental value for $N_c^{eff}=2 $, and the
CP violating parameter for this mode is quite small.

2. The branching ratio for the decay mode $B^- \to D^0 D^-$
lies below the present experimental upper bound and the
CP violating parameter for this mode is quite significant.
The number of charged $B$ mesons required to observe this
CP violating signal to three standard deviation is given as $N_B
=9/(BR \times a_{CP}^2) \approx 7.9 \times 10^6 $,
which is easily accessible with the currently running $ B $ factories.

It has been emphasized in Refs. [16-19] that the neutral B meson decay modes
to two charmed mesons can be used to measure the unitarity angle $\beta $
as alternative to the gold plated mode $B\to J/\psi K$. We argue further here
that the mode $B^- \to D^0 D^-$ can be used to quickly settledown the
search for observing direct CP violation
outside Kaon system, if SM description
of CP violation is correct or else could
provide us a clear indication of the
presence of new physics. It should be
noted here that the decay mode is flavor
self-tagging and hence experimentally favourable.
Furthermore, since the branching
ratio is $O(10^{-3})$ in our case,
which is larger than the other competitive mode like
$B \to \pi K$ where the branching ratio
is $O (10^{-5})$ \cite{pdg00} and the direct CP
violation parameter $a_{cp}$ is significantly
large i.e., we have obtained $a_{cp} $ around
4 \% as to that of $-$1.4 \% in ref \cite{akl}, this decay mode
is certainly a better candidate
to observe CP violation in the first round of B-factory
experiments.

To summarize, since the modes we consider are direct decays and not time
dependent, they may be observed in any experimental setting where
large number of $B$ mesons are produced. Apart from the SLAC and KEK
asymmetric $B$ factories these include CLEO and hadronic $B$ experiments
such as HERA-b, BTeV, Collider Detectors at Fermilab (CDF), D0 and CERN LHC-b
or high luminosity $Z$ factory. As we have used generalised factorization
approximation along with BSW model and Lepage and Brodsky  model for penguin
and annihilation contributions, we might have introduced certain uncertainties.
Nevertheless, since the branching ratio obtained for $B^- \to D^0 D_s^-$ matches
very well with the experimental value, we point out here that the decay mode
$B^- \to D^0 D^- $ may also be pursued further in the first round of B-factory experiments
(where it can  easily be accessible) to observe direct CP violation or to provide
us a hint for the presence of new physics.

\section{Acknowdgements}
AKG would like to thank Council of Scientific and Industrial Research,
Government of India, for financial support.

\begin{table}
\caption{Numerical value of the effective Wilson coefficients
$c_i^{eff}$ for $ b \to s $  and $b \to d $ transitions.}
\vspace {0.2 true in}
\begin{tabular}{|c|c|c|c|c|}
\hline
\multicolumn{1}{|c|}{} &
\multicolumn{1}{|c}{$ b \to  s $ }&
\multicolumn{1}{c|}{} &
\multicolumn{1}{|c}{$ b \to d $}&
\multicolumn{1}{c|}{}\\
\hline
\multicolumn{1}{|c|}{} &
\multicolumn{1}{|c|}{Time-like }&
\multicolumn{1}{|c|}{Space-like} &
\multicolumn{1}{|c}{Time-like}&
\multicolumn{1}{c|}{Space-like}\\
\hline
$c_1^{eff}$  & 1.168 & 1.168 & 1.168 & 1.168 \\
$c_2^{eff}$ & $-0.366 $& $-0.366 $& $-0.366 $& $-0.366$   \\
$c_3^{eff}$ & 0.0225+i0.0044
& $-(0.0096 +i 0.0003)$ &
0.0197+i 0.005&
$-(0.0123-i 0.0066)$\\
$c_4^{eff}$ & $ -(0.0456+i0.0133)$
& $(0.0505+i0.0009)$ &
$-(0.0373+0.015)$&
$(0.0586-i0.0199)$\\
$c_5^{eff}$ & 0.0132+i0.0044
& $-(0.0189+i0.0003) $ &
0.0104+i0.005&
$-(0.0216-i0.0066)$\\
$c_6^{eff}$ &  $-(0.0478+i0.0133)$
& $(0.0483+i0.0009) $ &
$-(0.0395+i0.015)$&
$(0.0564-i0.0199)$ \\
$c_7^{eff}/\alpha $  &$-(0.0282+i0.0363)$
&$ -(0.0303-i0.0018)$ &
$-(0.0119+i0.0398)$&
$-(0.0143+i0.0391)$\\
$c_8^{eff}/\alpha $ & 0.055 & 0.055& 0.055 & 0.055 \\
$c_9^{eff}/\alpha $  & $-(1.4252+i0.0363)$
& $-(1.4273-i0.0018) $ &
$-(1.4089+i0.0398)$&
$-(1.4113+i0.0391)$\\
$c_{10}^{eff}/\alpha $  & 0.481 & 0.481& 0.481 & 0.481 \\
\hline
\end{tabular}
\end{table}

\begin{table}
\begin{center}
\caption{Branching ratio and CP Asymmetry in \%
for $ B^- \to D^0 D_s^-, D^0 D^- $ decay modes. }
\vspace {0.2 true in}
\begin{tabular}{|c|c|c|c|c|c|}
\hline

\multicolumn{1}{|c|}{Decay modes}&
\multicolumn{1}{|r}{Branching}&
\multicolumn{1}{c}{Ratios}&
\multicolumn{1}{l|}{$(BR)$}&
\multicolumn{1}{|r}{$CP$}&
\multicolumn{1}{c|}{Asymmetry}\\
\hline
\multicolumn{1}{|c|}{}&
\multicolumn{1}{|c|}{$N_c^{eff}=2$}&
\multicolumn{1}{|c|}{$N_c^{eff}=3$} &
\multicolumn{1}{|c|}{Expt.}&
\multicolumn{1}{|c|}{$N_c^{eff}=2$}&
\multicolumn{1}{|c|}{$N_c^{eff}=3$}\\
\hline
&&&&&\\
$B^- \to D^0 D_s^- $& $1.29 \times 10^{-2} $ &
$1.46 \times 10^{-2}$ & $(1.3 \pm 0.4)\times 10^{-2}$ &
0.18 & 0.18 \\
&&&&&\\
$B^- \to D^0 D^- $& $8.72 \times 10^{-4} $ &
$9.86 \times 10^{-4}$ & $<6.7 \times 10^{-3}$ &
3.62 & 3.62 \\
\hline
\end{tabular}
\end{center}
\end{table}
    
\end{document}